\documentclass[secnumarabic, graphics,floatfix, nofootinbib,tightenlines,nobibnotes, aps, prl, 12pt]{revtex4-1}
\usepackage{graphicx}
\usepackage[english]{babel}
\usepackage[utf8]{inputenc}
\usepackage{amsmath,amssymb}
\usepackage{amsfonts}
\usepackage{multirow}
\usepackage{pstricks}
\usepackage{float}
\usepackage{subfigure}
\usepackage{color}
\usepackage{epsfig}
\usepackage{pst-tools}

\begin{document}

\title{Centrality dependence of multiplicity fluctuations from a hydrodynamical approach}

\author{Hong-Hao Ma$^{1}$}
\author{Kai Lin$^{2,3}$}
\author{Wei-Liang Qian$^{3,4,1}$}
\author{Bin Wang$^{1,5,6}$}

\affiliation{$^{1}$ School of Physical Science and Technology, Yangzhou University, 225002, Yangzhou, Jiangsu, China}
\affiliation{$^{2}$ Hubei Subsurface Multi-scale Imaging Key Laboratory, Institute of Geophysics and Geomatics, China University of Geosciences, 430074, Wuhan, Hubei, China}
\affiliation{$^{3}$ Escola de Engenharia de Lorena, Universidade de S\~ao Paulo, 12602-810, Lorena, SP, Brazil}
\affiliation{$^{4}$ Faculdade de Engenharia de Guaratinguet\'a, Universidade Estadual Paulista, 12516-410, Guaratinguet\'a, SP, Brazil}
\affiliation{$^{5}$ Collaborative Innovation Center of IFSA (CICIFSA), Shanghai Jiao Tong University, 200240, Shanghai, China}
\affiliation{$^{6}$ School of Aeronautics and Astronautics, Shanghai Jiao Tong University, 200240, Shanghai, China}

%\date{\today}
\date{Jan. 20, 2020}

\begin{abstract}

As one of the possible signals for the whereabouts of the critical point on the QCD phase diagram, recently, the multiplicity fluctuations in heavy-ion collisions have aroused much attention.
It is a crucial observable of the Beam Energy Scan program of the Relativistic Heavy Ion Collider.
In this work, we investigate the centrality dependence of the multiplicity fluctuations regarding the recent measurements from STAR Collaboration.
By employing a hydrodynamical approach, the present study is dedicated to the noncritical aspects of the phenomenon.
To be specific, in addition to the thermal fluctuations, finite volume corrections, and resonance decay at the freeze-out surface, the model is focused on the properties of the hydrodynamic expansion of the system and the event-by-event initial fluctuations.
It is understood that the real signal of the critical point can only be obtained after appropriately subtracting the background, the latter is investigated in the present work.
Besides the experimental data, our results are also compared to those of the hadronic resonance gas, as well as transport models.


\end{abstract}

\maketitle

\section{I. Introduction}

Described by the quantum field theory, the properties of a static system located in the vicinity of a fixed point are governed by its universality class, in terms of the respective critical exponents.
The latter dictates the scaling relations among different observables, as well as power-law divergences of relevant physical quantities, such as correlation length and particle fluctuations.
In fact, it has been speculated that the chiral phase transition of the QCD matter is of second order, related to the spontaneous symmetry breaking of the QCD vacuum.
However, the non-perturbative nature of QCD pose tremendous difficulties for the analytic approach in terms of the theory of the renormalization group.
Therefore, one has to resort to either numerical or phenomenological methods.
Lattice QCD simulations~\cite{lattice-01,lattice-02} have found a smooth crossover at vanishing baryon density and for the limit of large strange quark mass.
On the other hand, at finite chemical potential, one usually turns to model calculations.
A considerable amount of efforts~\cite{Halasz:1998qr, Berges:1998rc, Stephanov:1998dy, Schwarz:1999dj, Fodor:2004nz} indicates the occurrence of a first-order phase transition between the hadronic and quark-gluon plasma (QGP).
Therefore, a critical point is expected where the first-order phase transition line terminates.

The hot and dense matter created in the heavy-ion nuclear collision, however, is not a static system.
It evolves in time and may pass through the vicinity of the critical point.
Moreover, the freeze-out hypersurface, where the hadrons become mostly free particles and stream to the detector, is not necessarily close to the critical point.
Nonetheless, it is expected that resultant final state hadrons carry valuable information about the critical phenomenon.

On the theoretical side, many efforts have been devoted to the topic~\cite{qcd-phase-fluctuations-review-01,qcd-phase-fluctuations-review-02}.
As an exceedingly complicated problem, it might be governed simultaneously by distinct mechanisms.
Effective field theory at finite temperature has been employed to evaluate the moments of particle multiplicities.
It is speculated by the so called $\sigma$ model~\cite{qcd-phase-fluctuations-02,qcd-phase-fluctuations-03,qcd-phase-fluctuations-04} that higher moments are sensitive to the phase structure of the QCD matter.
To be specific, the normalized fourth-order cumulant of multiplicity distribution might be a non-monotonic function of collision energy. 
Regarding the fact that the system created in heavy-ion collision evolves rapidly in time, it is essential to take the dynamical effect of system evolution into account.
Tentatives along this train of thought~\cite{hydro-chiral-01} also lead to a variety of models.
For instance, the chiral fluid dynamics~\cite{hydro-chiral-04,hydro-chiral-07,hydro-chiral-08,hydro-chiral-09} divides the physics into two parts.
On the one hand, the quark degree of freedom is treated to be in an equilibrated heat bath, which evolves mostly in accordance with a hydrodynamical picture.
On the other hand, the chiral field is responsible for carrying the physics of symmetry spontaneous breaking, and the transition is triggered by the scalar density of the quark field. 
In a more recent version of the chiral model, a resultant Langevin equation is used to describe the chiral field.
Besides, another vital factor, namely, the critical slowing-down, is implemented by another approach, the Hydro+~\cite{hydro-chiral-sigma-01} model.
Here, the physics in question is that in the vicinity of the critical point, the time scale to achieve local equilibrium becomes comparable to that for global equilibrium.
There are other relevant features, which are essential even in the framework of conventional hydrodynamics, due to the existence of a critical point.
These include the modification to the equation of state (EoS), thermal~\cite{hydro-fluctuations-03}, as well as non-equilibrium~\cite{hydro-fluctuations-02}, fluctuations on the freeze-out surface, experimental uncertainties and cuts, in addition to other spurious contributions~\cite{qcd-phase-fluctuations-08,qcd-phase-fluctuations-09}.

On the experimental side, the ongoing Beam Energy Scan (BES) program~\cite{RHIC-star-bes-01, RHIC-star-bes-03, RHIC-star-bes-05,RHIC-star-mul-fluctuations-03,RHIC-star-mul-fluctuations-04} at the Relativistic Heavy Ion Collider (RHIC) is dedicated to exploring the phase diagram of QCD.
The program is to carry out Au+Au collisions ranging from 3.0 to 62.4 GeV.
Therefore, the properties of the critical point in question, if any, is likely to be captured by the data. 
The measurements, in turn, are aiming at the high baryon density region of the QCD matter.
To be specific, the measured quantities are those sensitive to the underlying physics while accessible experimentally.
The higher cumulants of conserved charges, as well as their combinations, are promising candidates recalling the above discussions.
As a matter of fact, measurements of multiplicity fluctuations were carried out in BES-I and have been further scheduled for BES-II programs, and have drawn much attention recently in the literature.

For the latter, the conditions for the conservation of net-charges are explicitly considered, and the effect was shown to be substantial.
In addition, resonance decay was shown to cause nonnegligible deviation from pure statistical distributions~\cite{statistical-model-03,statistical-model-04,statistical-model-07}.
For the most part, the obtained results~\cite{statistical-model-05,statistical-model-06,statistical-model-07,statistical-model-08,statistical-model-09,statistical-model-10} are manifestly consistent with the experimental data~\cite{RHIC-star-mul-fluctuations-01,RHIC-star-mul-fluctuations-02}.

The present work involves an effort to address the multiplicity fluctuations from a hydrodynamic viewpoint.
It is a further progress concerning a recent attempt~\cite{sph-bes-01} to evaluate multiplicity fluctuations using a hydrodynamical model.
Our approach does not explicitly involve the physics of critical phenomenon as mentioned above, but it is aimed to provide an estimation of the background signal from a mostly thermalized expanding system.
In fact, it is understood, a significant portion of the measured multiplicity fluctuation comes from the thermal fluctuations.
This has been confirmed from the calculations by using the Hadron Resonance Gas (HRG) models in the context of either grand canonical ensemble (GCE)~\cite{statistical-model-05,statistical-model-06,statistical-model-07} or canonical ensemble regarding conserved charges~\cite{statistical-model-03,statistical-model-04,statistical-model-08}.
The present model takes into account thermal fluctuations by using the formalism of GCE.
Also, volume correction, as well as resonance decay, are implemented in the code regarding hadron emission.
The hydrodynamic evolution is solved by using the Smoothed Particle Hydrodynamics (SPH) algorithm.
In our approach, every fluid element, denoted by an SPH particle, corresponds to a quantum GCE.
The information concerning the system expansion is recorded on the freeze-out surface.

The rest of the paper is organized as follows.
In the following section, we briefly review our hydrodynamical model, as well as thermodynamical fluctuations and resonance decay.
The results of numerical simulations are presented and discussed in Section III.
The last section is dedicated to concluding remarks.

\section{II. A hydrodynamic approach for multiplicity fluctuations}

In this section, we briefly describe the model employed in the present study.
The temporal expansion of the system is described by SPheRIO~\cite{sph-review-01}, a hydrodynamic code for an ideal relativistic fluid based on the SPH algorithm. 
In terms of discrete Lagrangian coordinates, individual fluid motion is mimicked, known as SPH particle. 
For the time being, we neglect any dissipative effects, and the Cooper-Frye sudden freeze-out is assumed for a constant temperature.
The latter, as discussed below, provides the baseline to evaluate the thermal fluctuations in the corresponding local rest frame. 
For the present study, the model parameters have been determined as to reproduce the experimental data regarding the particle spectra~\cite{sph-eos-02,sph-vn-04,sph-v2-02,sph-corr-ev-04,sph-eos-03,sph-vn-04,sph-corr-ev-06,sph-corr-ev-08,sph-corr-ev-09}.

On the freeze-out surface, every SPH particle is treated as a quantum GCE at a given temperature.
Therefore, a hydrodynamic event can be viewed as a collection of GCE ensembles represented by SPH particles.
As discussed in Ref.~\cite{sph-bes-01}, we do not explicitly incorporate global charge conservation at the freeze-out surface. 
It is noted that progress has been made very recently about implementing canonical or microcanonical systems in a hydrodynamical approach~\cite{hydro-fluctuations-04}.

In a hydrodynamic approach, each fluid element is considered to be in local equilibrium.
Usually, the volume in a static homogeneous system should be replaced by a time-like 3-surface $\sigma_\mu$.
In the case of the SPH method, the latter is further expressed in terms of SPH degrees of freedom.
To be specific,
\begin{eqnarray}
E\frac{d^3N_i}{dp^3}=\sum_j \frac{\nu_j n_{j\mu}p^{\mu}}{s_j|n_{j\rho}u_j^{\rho}|}\theta(u_{j\delta}p^{\delta})\langle n_{i}(u_{j\nu}p^{\nu}, x) \rangle, \label{d3n-sph}
\end{eqnarray}
where $p$ and $E$ are the momentum and energy of the emitted hadron, the subscript $i$ indicates particle species, the sum in $j$ is carried out for SPH particles, $\nu_j$ and $s_j$ represent the total entropy and entropy density of the $j$-th SPH particle.
$n_{i}(u_{j\nu}p^{\nu}, x)$ is defined in the context of a static statistical ensemble in the co-moving frame, as will be discussed below.
Subsequently, the ensemble average of particle number reads
\begin{eqnarray}
\langle N_{i} \rangle=\int p_{\bot}dp_{\bot}dy d\phi \sum_j \frac{\nu_j n_{j\mu}p^{\mu}}{s_j|n_{j\rho}u_j^{\rho}|}\theta(u_{j\delta}p^{\delta})\langle n_{i}(u_{j\nu}p^{\nu}, x) \rangle .\label{NavgSPH}
\end{eqnarray}

It is straightforward to show that the covariance is
\begin{eqnarray}
\langle \Delta N_{i}\Delta N_{j} \rangle= \int p_{\bot}dp_{\bot}dy d\phi \sum_j \frac{\nu_j n_{j\mu}p^{\mu}}{s_j|n_{j\rho}u_j^{\rho}|}\theta(u_{j\delta}p^{\delta}) v_{i}^2(u_{j\nu}p^{\nu}, x) ,\label{NcovSPH}
\end{eqnarray}
where, again, $v_{i}^2(u_{j\nu}p^{\nu}, x)$ is defined in the context of the static statistical ensemble.
The formalism for high order moments can be derived in a similar fashion~\cite{sph-bes-01}.

On an event-by-event basis, the hadrons emitted from an individual SPH particle, in its co-moving frame of reference, can be treated as a statistical ensemble.
For the latter, the particle number fluctuations have been extensively discussed, and relevant results can be found in Ref.~\cite{book-landau-5}.
To be specific, the GCE average value and variance of the occupation density in the momentum space read~\cite{statistical-model-03,statistical-model-04}
\begin{eqnarray}
\langle n_{p,i} \rangle= \frac{1}{\exp\left[(\sqrt{p^2+m_i^2}-\mu_i)/T\right]-\gamma_i} , 
\end{eqnarray}
\begin{eqnarray}
\langle \Delta n_{p,i}^2 \rangle \equiv  \langle (n_{p,i} - \langle n_{p,i} \rangle)^2 \rangle = \langle n_{p,i} \rangle (1+\gamma_i \langle n_{p,i} \rangle) ,\label{Delta_npi2}
\end{eqnarray}
where $T$ is the temperature, $m_i, \mu_i$ are the particle mass and chemical potential of species $i$, respectively, $\gamma_i$ corresponds to Bose ($+1$), Fermi ($-1$) or Boltzmann ($0$) statistics.

In our approach, the fluctuations are independent for different particle species as well as different momentum space, the covariance reads
\begin{eqnarray}
\langle \Delta n_{p,i} \Delta n_{k,j} \rangle = \delta_{ij}\delta_{pk}v_{p,i}^2 , \label{covnpi}
\end{eqnarray}
where $\Delta n_{p,i} = n_{p,i} - \langle n_{p,i}\rangle$, and $v_{p,i}^2=\langle \Delta n_{p,i}^2 \rangle$, defined in Eq.~(\ref{Delta_npi2}).

By summing up different momentum states, the average number of particles of species $i$ is found to be
\begin{eqnarray}
\langle N_{i} \rangle= \sum_p \langle n_{p,i} \rangle = \frac{g_i V}{2\pi^2}\int_0^{\infty} p^2 dp \langle n_{p,i} \rangle . \label{avgNi}
\end{eqnarray}

The variance and covariance can be evaluated as follows
\begin{eqnarray}
\langle \Delta N_{i}\Delta N_{j} \rangle= \sum_{p,k} \langle \Delta n_{p,i}\Delta n_{k,j} \rangle = \delta_{ij}\sum_p v_{p,i}^2 .\label{thermoCovariance}
\end{eqnarray}

Besides, higher statistical moments of multiplicity distributions like skewness $S\propto \langle \Delta N^3\rangle$ and kurtosis $\kappa \propto \langle \Delta N^4\rangle$ are also of particular importance.
These quantities can be evaluated, and the resultant expressions can be found in the Appendix of Ref.~\cite{sph-bes-01}.

The resonance decay can be considered by employing the generating function method introduced in Ref.~\cite{statistical-model-03}.
In general, resonance decay brings about additional fluctuations on top of the thermodynamical ones.

\section{III. Numerical results and Discussions}

\begin{figure}[htb]
\begin{tabular}{cc}
\begin{minipage}{300pt}
\centerline{\includegraphics[width=300pt]{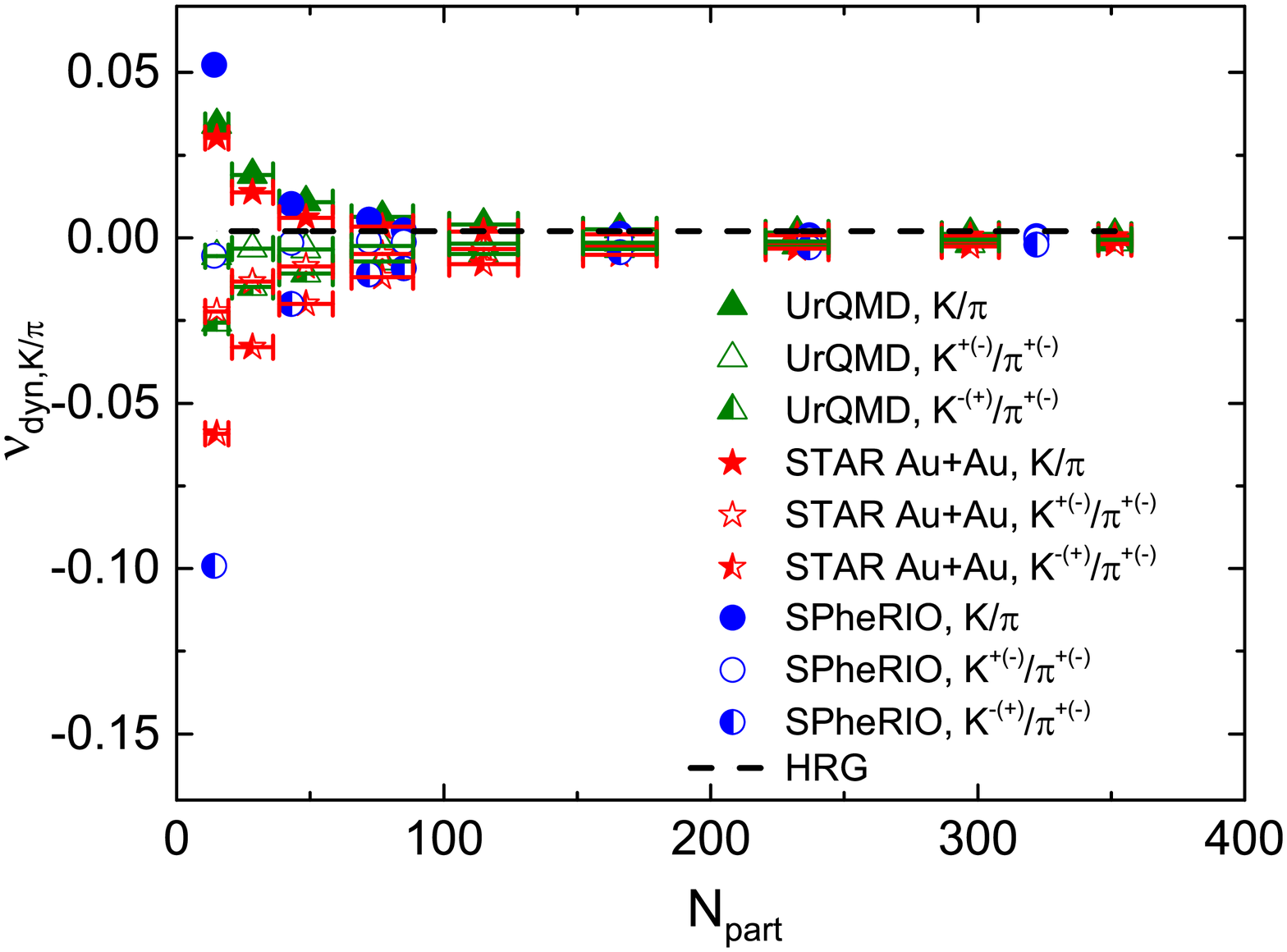}}
\end{minipage}
\\
\begin{minipage}{300pt}
\centerline{\includegraphics[width=300pt]{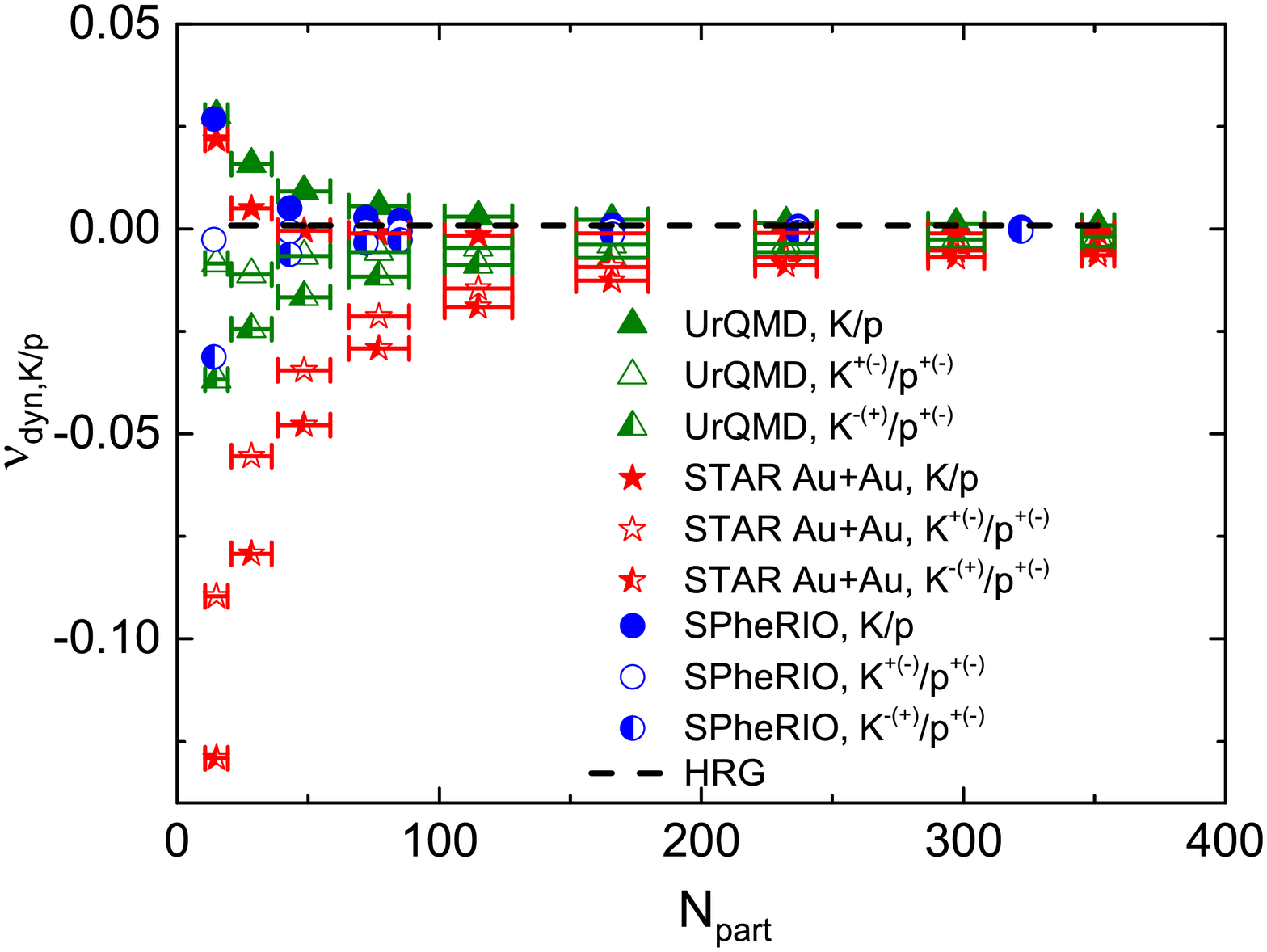}}
\end{minipage}
\\
\begin{minipage}{300pt}
\centerline{\includegraphics[width=300pt]{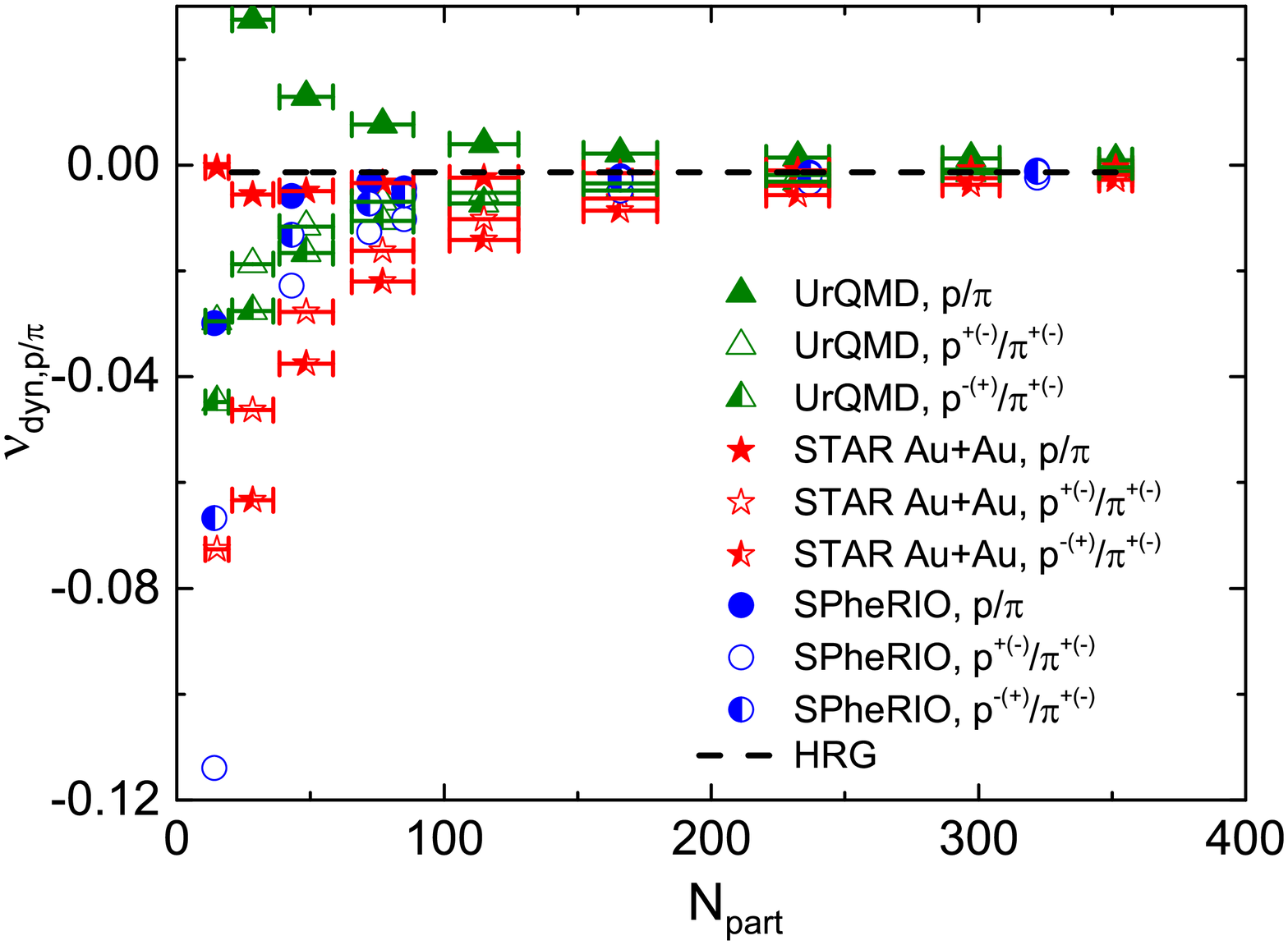}}
\end{minipage}
\end{tabular}
\caption{The centrality dependence of the calculated dynamical fluctuations of particle ratio $K/ \pi$, $p/ \pi$ and $K/p$ for Au+Au collisions at 200GeV. The STAR data are from Ref.~\cite{RHIC-star-mul-fluctuations-03}, presented by filled, open, half-filled red stars. 
The SPheRIO results are given by the filled, open, half-filled blue circle.
The UrQMD model calculations are shown in the filled, open, half-filled green triangle.
The dashed black line presents the HRG calculations.
 }\label{nudyn}
\end{figure}

We carried out hydrodynamic simulations of Au+Au collisions at 200 GeV by using SPheRIO code for different centrality windows.
The number of participants of each event is thus subtracted, and the events are thus reorganized in accordance with the existing data of the BES program~\cite{RHIC-star-mul-fluctuations-03, RHIC-star-mul-fluctuations-04}.

The IC are generated by using NeXuS~\cite{nexus-1,nexus-rept}.
The results presented below are from simulations carried out for 980 events for the 0 - 5\%, 1039 events for the 6\% - 15\%, 1138 events for the 15\% - 25\%, 1112 events for the 25\% - 35\%, 994 events for the 35\% - 45\%, 966 events for the 45\% - 55\%, and 985 events for the 60\% - 80\% centrality windows, respectively.
The number of participants is extracted from individual events and then assigned to respective centrality bins.
For a given bin, the average number of participants, as well as the IC, are calculated.
Subsequently, the hydrodynamical calculations are carried out by employing the averaged IC.

In Fig.~\ref{nudyn}, we show the calculated dynamical fluctuations of particle ratios $K/\pi$, $K/p$, and $p/\pi$ as functions of the numbers of participants. 
By definition, the quantity $\nu_{\rm dyn,p/\pi}$ measures the deviation in the ratios of $p/\pi$ from those of a Poissonian distribution.
To be specific, it reads
\begin{eqnarray}
\nu_{\rm dyn,p/\pi} = \frac{\langle N_{p}(N_{p}-1)\rangle}{\langle N_{p}\rangle^{2}} + \frac{\langle N_{\pi}(N_{\pi}-1)\rangle}{\langle N_{\pi}\rangle^{2}} - 2\frac{\langle N_{p}N_{\pi}\rangle}{\langle N_{p}\rangle\langle N_{\pi}\rangle}\ .
\label{nudyn}
\end{eqnarray}
The results of hydrodynamic simulations by SPheRIO, and those of UrQMD as well as HRG models are presented together with the data from the STAR Collaborations~\cite{RHIC-star-mul-fluctuations-03}.

The SPheRIO results show a reasonably consistent trend observed in the data.
A similar agreement was also obtained by the transport model, UrQMD.
These two approaches are intrinsically different from that for a static system, in terms of the HRG model with resonance decays.
The latter does not depend on the volume or overall multiplicity. 
On the other hand, it is observed that the results from SPheRIO or UrQMD do scale with multiplicity.
This indicates that the difference can be attributed to the finite volume of the system, associated with the system expansion, as presented in both models.
For more central collisions, the obtained $N_{\mathrm{part}}$ dependence of dynamical fluctuations from SPheRIO and UrQMD calculations are less prominent, and the magnitudes are close to that of HRG model.
As one goes to more peripheral collisions, the centrality dependence becomes more pronounced, especially when $N_{\mathrm{part}} < 100$.
For the hydrodynamical viewpoint, the observed difference from a pure thermal ensemble is due to the inhomogenous and anisotropic freeze-out surface elements, which is originated from the IC.
For more peripheral collisions, the IC becomes more irregular as the hydrodynamical picture starts to fail.
For event-by-event initial conditions, it has been observed~\cite{sph-bes-01} that the multiplicity fluctuations becomes the dominant factor.
Therefore, from a theoretical viewpoint, the present framework provides a possibility to separate several different sources of multiplicity fluctuations, namely, the thermal fluctuations, resonance decay, hydrodynamical expansion, and initial state event-by-event fluctuations.
As anyone of the above factors can be switched on and off independent of others, their respective effect can be investigated individually.

\begin{figure}[htb]
\begin{tabular}{cc}
\begin{minipage}{300pt}
\centerline{\includegraphics[width=300pt]{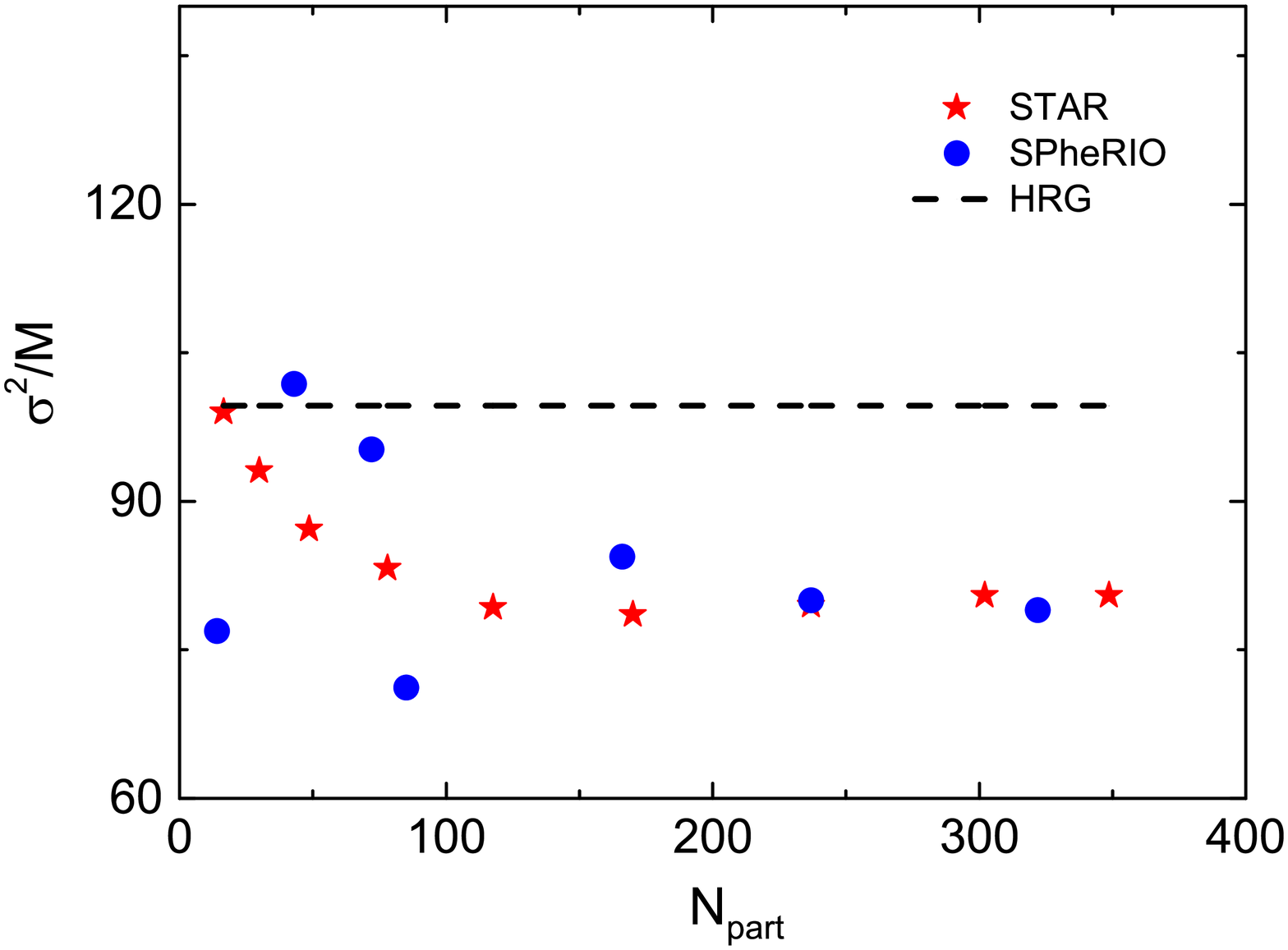}}
\end{minipage}
\\
\begin{minipage}{300pt}
\centerline{\includegraphics[width=300pt]{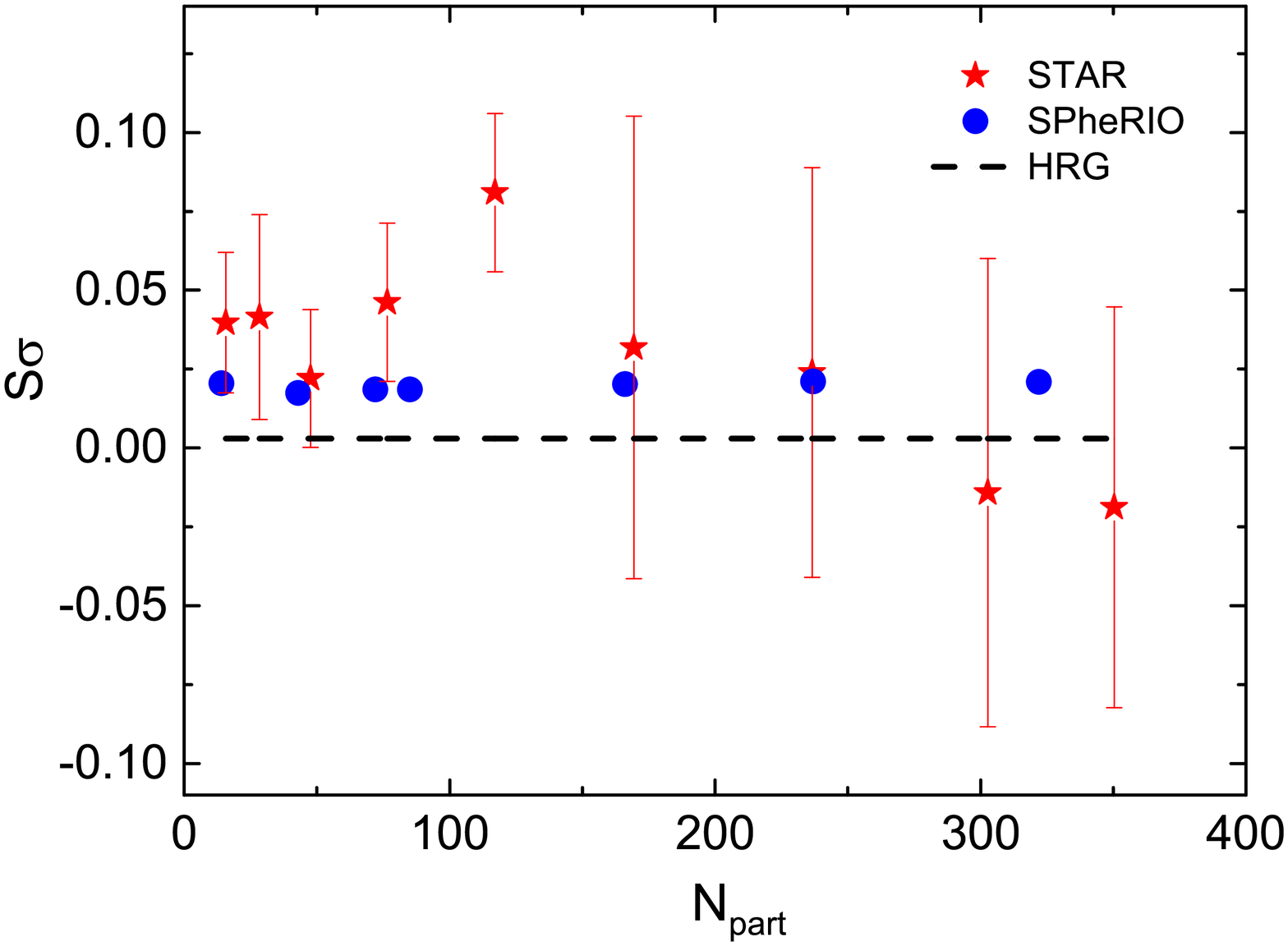}}
\end{minipage}
\\
\begin{minipage}{300pt}
\centerline{\includegraphics[width=300pt]{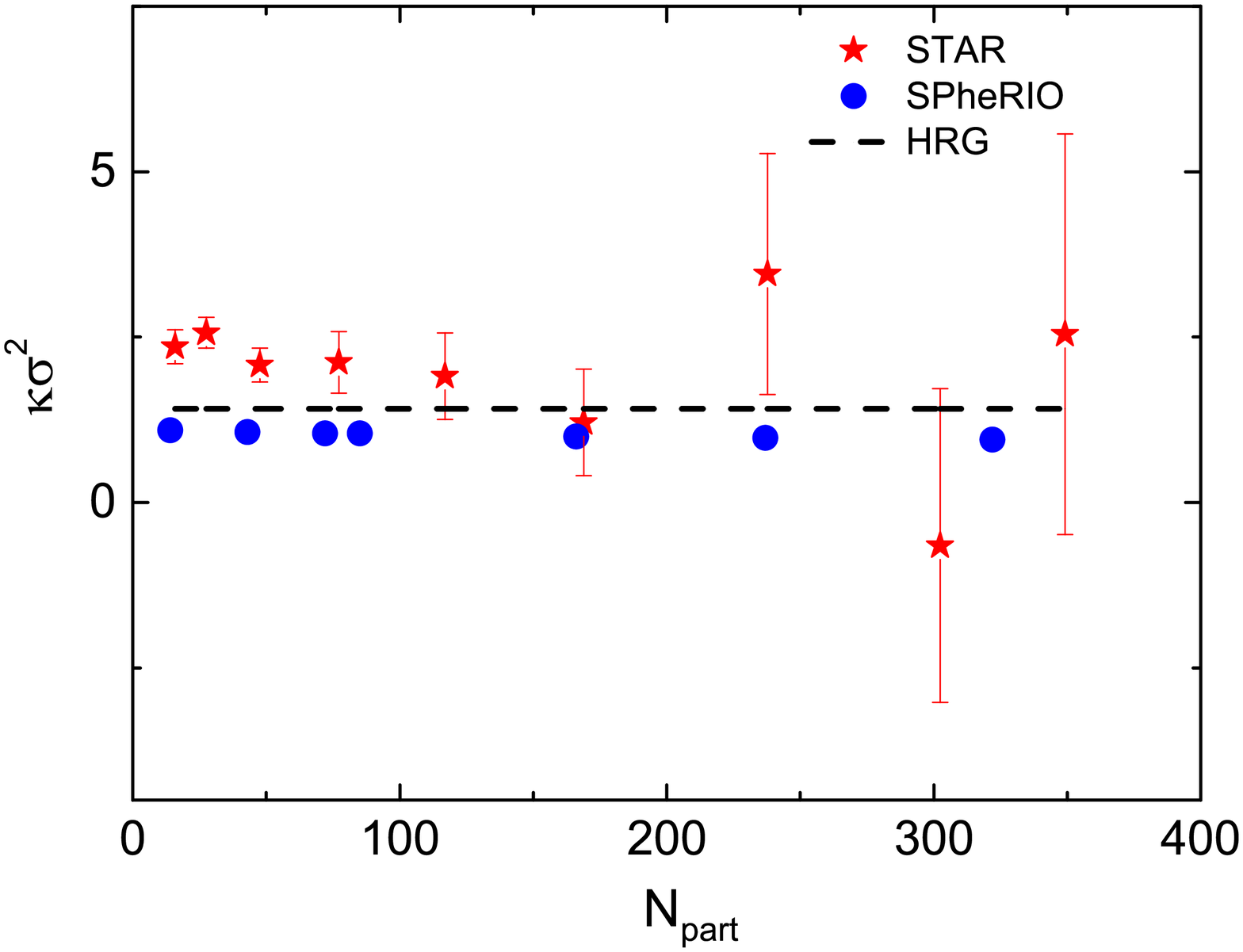}}
\end{minipage}
\end{tabular}
\caption{(Color online) The centrality dependence of the calculated higher moments of multiplicity of net-charge.
The results are for Au+Au collisions at 200GeV. 
The STAR data are from Ref.~\cite{RHIC-star-mul-fluctuations-04}, presented by filled red stars. 
The SPheRIO results are shown in the filled blue circle, and the dashed black line gives that from the HRG calculations.
}\label{fpratios}
\end{figure}

In Fig.~\ref{fpratios}, we present various cumulant ratios at different centrality obtained by SPheRIO.
The corresponding results obtained by the HRG models are also shown.
%The SPheRIO results are those of averaged IC.
The STAR measurements~\cite{RHIC-star-mul-fluctuations-02} are for Au+Au collisions at $200$ GeV.
The products $\kappa\sigma^2$ and $S\sigma$ are determined in terms of the ratios of particle number cumulants, which can be evaluated by using their respective definitions and the information of the freeze-out surface.
In particular, $\kappa\sigma^2$ is expected to be equal to 1 for ideal Poissonian distribution.
Therefore, the calculated quantity measures the deviation from a static homogeneous classical ensemble.
For a hydrodynamic approach, such deviations come from the inhomogeneity, collective motion, resonance decay, and event-by-event IC fluctuations. 
Numerically, the results by the HRG model indicate that resonance decay does not imply a significant effect, when comparing against other factors.
SPheRIO results present a similar tendency for the cases of $\sigma^2/M$ and $S\sigma$, when compared to those of STAR data.
This indicates a significant part of the net-charge fluctuations can be understood within the framework of an approach that appropriately considers the system expansion.
On the other hand, the resultant $\kappa\sigma^2$, although nonvanishing, are found to be relatively insignificant with respect to those of $\sigma^2/M$ and $S\sigma$.
It is particularly evident when one compares those against the difference between the experimental data and HRG results.
By definition, kurtosis involves contributions up to the fourth moment of the particle distribution.
Compared to the other two quantities, it measures higher moments of multiplicity fluctuations and is potentially more sensitive to the critical phenomenon~\cite{qcd-phase-fluctuations-02,qcd-phase-fluctuations-03}.
While the hydrodynamical model has mostly captured the characteristics of $\sigma^2/M$ and $S\sigma$, the same approach is shown to be incapable of reproducing the main feature of the measured $\kappa\sigma^2$.
The hydrodynamic results on kurtosis do not show a significant difference from those of HRG.
In fact, the effect of system expansion seems to push the results towards a ``wrong" direction further.
Therefore, it is speculated that the experimentally observed non-monotonical dependence of kurtosis on the centrality might imply crucial information, as it might be an interesting observable regarding the goal to capture crucial information about the critical point.
Unfortunately, it lies out of the scope of the present approach.

\section{IV. Concluding remarks}

To summarize, this work is devoted to studying the centrality dependence of the multiplicity fluctuations in heavy-ion nuclear collisions by using the hydrodynamical model SPheRIO.
Our approach is focused on some of the noncritical aspects of the multiplicity fluctuations.
This is because apart from the critical phenomenon in the vicinity of the critical point, many other possible sources also affect the multiplicity fluctuations.
To extract relevant pieces of information from the experimental data, these non-critical fluctuations need to be carefully subtracted.
Previous studies by the HRG model, the effects of thermal fluctuations, finite volume correction, and resonance decay have been studied.   
The present model continues the ongoing effort to include further additional features encoded in terms of the freeze-out process. 
To be specific, we further investigate the effect of the IC, as well as the temporal expansion of the system.
The hydrodynamical model employed in the study does not contain any additional free parameters, as the existing ones have been determined in reproducing the particle spectra in previous studies.

The obtained results are compared to those of the HRG, UrQMD models, as well as the experimental data.
Overall, regarding the existing data, the results obtained by SPheRIO shows that the hydrodynamical expansion plays an essential role regarding the centrality dependence of multiplicity fluctuations.
From the theoretical viewpoint, the calculations of centrality dependence might provide a scheme to separate the effects of various non-critical aspects concerning multiplicity fluctuations.
It might be meaningful to carry out a further regarding other aspects, such as the effect of EoS, particularly regarding the region with finite baryon density. 
Further study in this direction is in progress.   

\section*{Acknowledgments}
We are thankful for valuable discussions with Nu Xu, Fr\'ed\'erique Grassi, De-Chang Dai, and Matthew Luzum. 
We gratefully acknowledge the financial support from
Funda\c{c}\~ao de Amparo \`a Pesquisa do Estado de S\~ao Paulo (FAPESP),
Funda\c{c}\~ao de Amparo \`a Pesquisa do Estado do Rio de Janeiro (FAPERJ),
Conselho Nacional de Desenvolvimento Cient\'{\i}fico e Tecnol\'ogico (CNPq),
and Coordena\c{c}\~ao de Aperfei\c{c}oamento de Pessoal de N\'ivel Superior (CAPES).
A part of the work was developed under the project INCTFNA Proc. No. 464898/2014-5.
This research is also supported by the Center for Scientific Computing (NCC/GridUNESP) of the S\~ao Paulo State University (UNESP).

\bibliographystyle{h-physrev-qian}
\bibliography{references_qian,references_ma}

\begin{thebibliography}{10}

\bibitem{lattice-01}
Z.~Fodor and S.~Katz,
\newblock JHEP {\bf 0203} (2002) 014, arXiv:hep-lat/0106002.

\bibitem{lattice-02}
F.~Karsch,
\newblock Nucl.Phys. {\bf A698} (2002) 199, arXiv:hep-ph/0103314.

\bibitem{Halasz:1998qr}
A.~M. Halasz, A.~D. Jackson, R.~E. Shrock, M.~A. Stephanov, and J.~J.~M.
  Verbaarschot,
\newblock Phys. Rev. {\bf D58} (1998) 096007, arXiv:hep-ph/9804290.

\bibitem{Berges:1998rc}
J.~Berges and K.~Rajagopal,
\newblock Nucl. Phys. {\bf B538} (1999) 215, arXiv:hep-ph/9804233.

\bibitem{Stephanov:1998dy}
M.~A. Stephanov, K.~Rajagopal, and E.~V. Shuryak,
\newblock Phys. Rev. Lett. {\bf 81} (1998) 4816, arXiv:hep-ph/9806219.

\bibitem{Schwarz:1999dj}
T.~M. Schwarz, S.~P. Klevansky, and G.~Papp,
\newblock Phys. Rev. {\bf C60} (1999) 055205, arXiv:nucl-th/9903048.

\bibitem{Fodor:2004nz}
Z.~Fodor and S.~D. Katz,
\newblock JHEP {\bf 04} (2004) 050, arXiv:hep-lat/0402006.

\bibitem{qcd-phase-fluctuations-review-01}
M.~Asakawa and M.~Kitazawa,
\newblock Prog. Part. Nucl. Phys. {\bf 90} (2016) 299, arXiv:1512.05038.

\bibitem{qcd-phase-fluctuations-review-02}
X.~Luo and N.~Xu,
\newblock Nucl. Sci. Tech. {\bf 28} (2017) 112, arXiv:1701.02105.

\bibitem{qcd-phase-fluctuations-02}
M.~A. Stephanov,
\newblock Phys. Rev. Lett. {\bf 102} (2009) 032301, arXiv:0809.3450.

\bibitem{qcd-phase-fluctuations-03}
M.~A. Stephanov,
\newblock Phys. Rev. Lett. {\bf 107} (2011) 052301, arXiv:1104.1627.

\bibitem{qcd-phase-fluctuations-04}
B.~J. Schaefer and M.~Wagner,
\newblock Phys. Rev. {\bf D85} (2012) 034027, arXiv:1111.6871.

\bibitem{hydro-chiral-01}
K.~Paech, H.~Stoecker, and A.~Dumitru,
\newblock Phys. Rev. {\bf C68} (2003) 044907, arXiv:nucl-th/0302013.

\bibitem{hydro-chiral-04}
D.~T. Son and P.~Surowka,
\newblock Phys. Rev. Lett. {\bf 103} (2009) 191601, arXiv:0906.5044.

\bibitem{hydro-chiral-07}
M.~Nahrgang, S.~Leupold, C.~Herold, and M.~Bleicher,
\newblock Phys. Rev. {\bf C84} (2011) 024912, arXiv:1105.0622.

\bibitem{hydro-chiral-08}
M.~Nahrgang, C.~Herold, S.~Leupold, I.~Mishustin, and M.~Bleicher,
\newblock J. Phys. {\bf G40} (2013) 055108, arXiv:1105.1962.

\bibitem{hydro-chiral-09}
C.~Herold, M.~Nahrgang, I.~Mishustin, and M.~Bleicher,
\newblock Phys. Rev. {\bf C87} (2013) 014907, arXiv:1301.1214.

\bibitem{hydro-chiral-sigma-01}
M.~Stephanov and Y.~Yin,
\newblock Phys. Rev. {\bf D98} (2018) 036006, arXiv:1712.10305.

\bibitem{hydro-fluctuations-03}
J.~Li, H.-j. Xu, and H.~Song,
\newblock Phys. Rev. {\bf C97} (2018) 014902, arXiv:1707.09742.

\bibitem{hydro-fluctuations-02}
L.~Jiang, P.~Li, and H.~Song,
\newblock Phys. Rev. {\bf C94} (2016) 024918, arXiv:1512.06164.

\bibitem{qcd-phase-fluctuations-08}
M.~Hippert, E.~S. Fraga, and E.~M. Santos,
\newblock Phys. Rev. {\bf D93} (2016) 014029, arXiv:1507.04764,
\newblock [Phys. Rev.D93,014029(2016)].

\bibitem{qcd-phase-fluctuations-09}
M.~Hippert and E.~S. Fraga,
\newblock Phys. Rev. {\bf D96} (2017) 034011, arXiv:1702.02028.

\bibitem{RHIC-star-bes-01}
STAR, B.~Mohanty,
\newblock J. Phys. {\bf G38} (2011) 124023, arXiv:1106.5902.

\bibitem{RHIC-star-bes-03}
STAR, L.~Kumar,
\newblock Nucl. Phys. {\bf A904-905} (2013) 256c, arXiv:1211.1350.

\bibitem{RHIC-star-bes-05}
STAR, C.~Yang,
\newblock Nucl. Phys. {\bf A967} (2017) 800.

\bibitem{RHIC-star-mul-fluctuations-03}
H.~Wang,
\newblock {\em {Study of particle ratio fluctuations and charge balance
  functions at RHIC}},
\newblock PhD thesis, Michigan State U., 2012, 1304.2073.

\bibitem{RHIC-star-mul-fluctuations-04}
STAR, L.~Adamczyk {\em et~al.},
\newblock Phys. Rev. Lett. {\bf 113} (2014) 092301, arXiv:1402.1558.

\bibitem{statistical-model-03}
V.~Begun, M.~I. Gorenstein, M.~Hauer, V.~Konchakovski, and O.~Zozulya,
\newblock Phys.Rev. {\bf C74} (2006) 044903, arXiv:nucl-th/0606036.

\bibitem{statistical-model-04}
F.~Becattini, A.~Keranen, L.~Ferroni, and T.~Gabbriellini,
\newblock Phys. Rev. {\bf C72} (2005) 064904, arXiv:nucl-th/0507039.

\bibitem{statistical-model-07}
J.~Fu,
\newblock Phys. Lett. {\bf B722} (2013) 144.

\bibitem{statistical-model-05}
J.~Fu,
\newblock Phys. Lett. {\bf B679} (2009) 209.

\bibitem{statistical-model-06}
J.~Fu,
\newblock Phys. Rev. {\bf C85} (2012) 064905.

\bibitem{statistical-model-08}
J.-H. Fu,
\newblock Phys. Rev. {\bf C96} (2017) 034905, arXiv:1610.07138.

\bibitem{statistical-model-09}
F.~Karsch and K.~Redlich,
\newblock Phys. Lett. {\bf B695} (2011) 136, arXiv:1007.2581.

\bibitem{statistical-model-10}
P.~Garg {\em et~al.},
\newblock Phys. Lett. {\bf B726} (2013) 691, arXiv:1304.7133.

\bibitem{RHIC-star-mul-fluctuations-01}
STAR, T.~J. Tarnowsky,
\newblock Acta Phys. Polon. Supp. {\bf 5} (2012) 515, arXiv:1201.3336.

\bibitem{RHIC-star-mul-fluctuations-02}
STAR, J.~Thader,
\newblock Nucl. Phys. {\bf A956} (2016) 320, arXiv:1601.00951.

\bibitem{sph-bes-01}
H.-H. Ma {\em et~al.},
\newblock (2019), arXiv:1910.00705.

\bibitem{sph-review-01}
Y.~Hama, T.~Kodama, and O.~Socolowski~Jr.,
\newblock Braz. J. Phys. {\bf 35} (2005) 24, arXiv:hep-ph/0407264.

\bibitem{sph-eos-02}
W.-L. Qian {\em et~al.},
\newblock Braz. J. Phys. {\bf 37} (2007) 767, arXiv:nucl-th/0612061.

\bibitem{sph-vn-04}
W.-L. Qian {\em et~al.},
\newblock J.Phys.G {\bf G41} (2014) 015103, arXiv:1305.4673.

\bibitem{sph-v2-02}
R.~Andrade, F.~Grassi, Y.~Hama, T.~Kodama, and W.~Qian,
\newblock Phys.Rev.Lett. {\bf 101} (2008) 112301, arXiv:0805.0018.

\bibitem{sph-corr-ev-04}
W.-L. Qian, R.~Andrade, F.~Gardim, F.~Grassi, and Y.~Hama,
\newblock Phys.Rev. {\bf C87} (2013) 014904, arXiv:1207.6415.

\bibitem{sph-eos-03}
D.~M. Dudek {\em et~al.},
\newblock Int. J. Mod. Phys. {\bf E27} (2018) 1850058, arXiv:1409.0278.

\bibitem{sph-corr-ev-06}
W.~M. Castilho, W.-L. Qian, F.~G. Gardim, Y.~Hama, and T.~Kodama,
\newblock Phys.Rev. {\bf C95} (2017) 064908, arXiv:1610.04108.

\bibitem{sph-corr-ev-08}
W.~M. Castilho, W.-L. Qian, Y.~Hama, and T.~Kodama,
\newblock Phys. Lett. {\bf B777} (2018) 369, arXiv:1707.09878.

\bibitem{sph-corr-ev-09}
W.~M. Castilho and W.-L. Qian,
\newblock Nucl. Phys. {\bf A974} (2018) 35, arXiv:1803.08903.

\bibitem{hydro-fluctuations-04}
D.~Oliinychenko and V.~Koch,
\newblock Phys. Rev. Lett. {\bf 123} (2019) 182302, arXiv:1902.09775.

\bibitem{book-landau-5}
L.~Landau and E.~Lifshitz,
\newblock {\em Statistical Physics, Part I}, Course of Theoretical Physics
  Vol.~5, 3 ed. (Butterworth-Heinemann, 1980).

\bibitem{nexus-1}
H.~Drescher, S.~Ostapchenko, T.~Pierog, and K.~Werner,
\newblock Phys.Rev. {\bf C65} (2002) 054902, arXiv:hep-ph/0011219.

\bibitem{nexus-rept}
H.~Drescher, M.~Hladik, S.~Ostapchenko, T.~Pierog, and K.~Werner,
\newblock Phys.Rept. {\bf 350} (2001) 93, arXiv:hep-ph/0007198.

\end{thebibliography}

\end{document}